\documentclass[prl,twocolumn,showpacs,amsmath,amssymb,superscriptaddress]{revtex4}
\usepackage{graphicx}
\usepackage{dcolumn}
\usepackage{bm}
\usepackage{epsfig}

\begin{document}


\title{Magnetism in the high-$T_{\mathrm{c}}$ analogue Cs$_{2}$AgF$_{4}$
studied with muon-spin relaxation}

\author{T. Lancaster}
\email{t.lancaster1@physics.ox.ac.uk} 
\author{S.J. Blundell} 
\author{P.J. Baker}
\author{W. Hayes}
\affiliation{Clarendon Laboratory, Oxford University Department of Physics, Parks
Road, Oxford, OX1 3PU, UK
}

\author{S.R. Giblin}
\author{S.E. McLain}
\author{F.L. Pratt}
\affiliation{ISIS Facility, Rutherford Appleton Laboratory, Chilton, Oxfordshire OX11 0QX, UK}

\author{Z. Salman}
\affiliation{Clarendon Laboratory, Oxford University Department of Physics, Parks
Road, Oxford, OX1 3PU, UK
}
\affiliation{ISIS Facility, Rutherford Appleton Laboratory, Chilton, Oxfordshire
OX11 0QX, UK}

\author{E.A. Jacobs}
\author{J.F.C. Turner}
\author{T. Barnes}
\affiliation{Department of Chemistry and Neutron Sciences Consortium,
University of Tennessee, Knoxville, Tennessee 37996, USA}
\date{\today}

\begin{abstract}
We present the results of a muon-spin relaxation study of the
high-$T_{\mathrm{c}}$ analogue material
Cs$_{2}$AgF$_{4}$.
We find unambiguous evidence for magnetic order, intrinsic to the
material, below $T_{\mathrm{C}}=13.95(3)$~K. 
The ratio of inter- to intraplane coupling is estimated to be
$|J'/J|=1.9 \times 10^{-2}$, while fits of the temperature
dependence of the order parameter reveal a critical exponent 
$\beta=0.292(3)$, implying
an intermediate character between pure two- and three- dimensional
magnetism in the critical regime. 
Above $T_{\mathrm{C}}$ we observe
a signal characteristic of dipolar interactions due to
 linear F--$\mu^{+}$--F bonds, allowing the  muon stopping
sites in this compound to be characterized.
\end{abstract}
\maketitle

\begin{figure}
\begin{center}
\epsfig{file=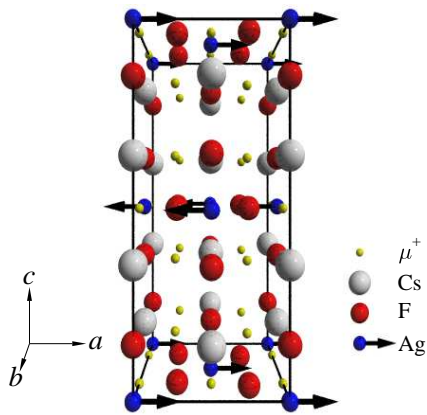,width=5.7cm}
\caption{(Color online.)
Structure of Cs$_{2}$AgF$_{4}$ showing a possible magnetic structure. 
Candidate muon sites occur in both the [CsF] and [AgF$_{2}$]
planes. 
\label{structure}}
\end{center}
\end{figure}

Twenty years after its discovery, high-$T_{\mathrm{c}}$
superconductivity remains one of the most pressing problems in
condensed matter physics. High-$T_{\mathrm{C}}$ cuprates share a layered
structure of [CuO$_{2}$] planes with strong antiferromagnetic (AFM)
interactions between $S=1/2$ 3$d^{9}$ Cu$^{2+}$ ions \cite{bg1,bg2}. 
However, analogous
materials based upon 3$d$ transition metal systems 
such as manganites \cite{dagotto}
and nickelates \cite{cava} share neither the magnetic nor the superconducting
properties of the high-$T_{\mathrm{C}}$ cuprates, leading to speculation that the
spin-$\frac{1}{2}$ character of Cu$^{2+}$ is unique in this context. 
A natural extension to this line of inquiry is to explore compounds 
based on the 4$d$ analogue of Cu$^{2+}$, namely $S=\frac{1}{2}$ 4$d^{9}$
Ag$^{2+}$ \cite{grochala};
 this motivated the synthesis of the
layered fluoride Cs$_{2}$AgF$_{4}$ which contains silver in the
unusual divalent oxidation state \cite{odenthal,mclain}. 
This material possesses several structural similarities with the
superconducting parent compound La$_{2}$CuO$_{4}$; it is comprised of planes
of [AgF$_{2}$] instead of [CuO$_{2}$] separated by planes of [CsF]
instead of [LaO]
(Fig.~\ref{structure}).

Magnetic measurements \cite{mclain} suggest that
in contrast to the antiferromagnetism
of La$_{2}$CuO$_{4}$, Cs$_{2}$AgF$_{4}$ is well modelled as a 
two-dimensional (2D) Heisenberg ferromagnet 
(described by the Hamiltonian 
$\mathcal{H}=J \sum_{\langle i j \rangle} \mathbf{S}_{i} \cdot \mathbf{S}_{j}$)
with intralayer coupling $J/k_{\mathrm{B}}=44.0$~K. 
The observation of a magnetic transition below $T_{\mathrm{C}}\approx
15$~K with no spontaneous magnetization
in zero applied field (ZF) and a small saturation magnetization
($\sim 40$~mT), suggests the existence of a weak, AFM
interlayer coupling. 
This behavior is reminiscent of the 2D ferromagnet K$_{2}$CuF$_{4}$
\cite{yamada}, where ferromagnetic (FM) exchange results from orbital
ordering driven by a Jahn-Teller distortion
\cite{ito,khomskii}.
On this basis, it has been suggested  that in Cs$_{2}$AgF$_{4}$
a staggered ordering of
$d_{z^{2}-x^{2}}$ and $d_{z^{2}-y^{2}}$ hole-containing orbitals on
the Ag$^{2+}$ ions gives rise to the FM superexchange \cite{mclain}.
An alternative scenario has also been advanced  on the
basis of density functional calculations in which a 
$d_{3z^{2}-r^{2}}-p-d_{x^{2}-y^{2}}$
orbital interaction through the Ag--F--Ag bridges causes spin
polarization of the $d_{x^{2}-y^{2}}$ band \cite{dai}. 

Although inelastic neutron scattering measurements have been carried 
out on this 
material \cite{mclain},
 both Cs and Ag strongly
absorb neutrons, 
resulting in limited resolution and a poor signal-to-noise ratio.
In contrast,
spin-polarized muons, which are very sensitive probes of local magnetic
fields, suffer no such impediments and, as we shall see, are ideally 
suited to investigations of the magnetism in fluoride materials.

In this paper we present the results of a ZF muon-spin relaxation 
($\mu^{+}$SR) investigation of Cs$_{2}$AgF$_{4}$. We are able to
confirm that the material is uniformly ordered throughout its bulk
below $T_{\mathrm{C}}$ and show that the critical behavior associated
with the magnetic phase transition is intermediate in character between 2D and 
3D. 
In addition, strong coupling between the muon and F$^{-}$ ions allows us to
characterise the muon stopping states in this compound.

ZF $\mu^{+}$SR measurements were
made on the MuSR instrument at the ISIS facility,
using an Oxford Instruments Variox $^4$He cryostat.
In a $\mu^{+}$SR experiment spin-polarized
positive muons are stopped in a target sample, where the muon usually
occupies an interstitial position in the crystal. 
The observed property in the experiment is the time evolution of the
muon spin polarization, the behavior of which depends on the
local magnetic field $B$ at the muon site \cite{steve}.
Polycrystalline Cs$_{2}$AgF$_{4}$ was synthesised as previously 
reported \cite{mclain}. 
Due to its chemical reactivity, the sample was mounted under an Ar atmosphere 
in a gold plated Ti sample holder with a cylindrical sample space of diameter 
2.5~cm  and depth 2~mm. 
A 25~$\mu$m thick window was screw-clamped onto a gold o-ring on the
main body of the sample holder resulting in an airtight seal.

\begin{figure}
\begin{center}
\epsfig{file=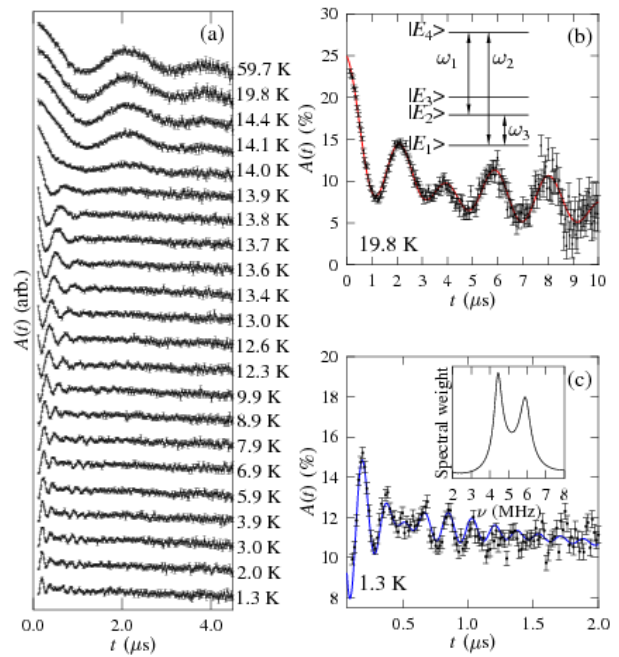,width=8cm}
\caption{ (Color online.)
(a) Temperature evolution of ZF $\mu^{+}$SR spectra measured
on Cs$_{2}$AgF$_{4}$ between 1.3~K and 59.7~K. (b) Above
$T_{\mathrm{C}}$ low frequency oscillations are observed due to the
dipole-dipole coupling of F--$\mu^{+}$--F states. {\it Inset:}
The energy level structure allows three transitions, leading to
three observed frequencies. (c) Below $T_{\mathrm{C}}$
higher frequency oscillations are observed
due to quasistatic magnetic fields at the muon sites. 
{\it Inset:} Maximum entropy
analysis reveal two magnetic frequencies corresponding
to two magnetically inequivalent muon sites.\label{data}}
\end{center}
\end{figure}

Example ZF $\mu^{+}$SR spectra measured on Cs$_{2}$AgF$_{4}$ are shown
in Fig.~\ref{data}(a). Below $T_{\mathrm{C}}$ (Fig.~\ref{data}(c))
we observe oscillations in the time dependence of the muon
polarization (the ``asymmetry'' $A(t)$ \cite{steve}) which are
characteristic of a quasi-static local magnetic field at the 
muon stopping site. This local field causes a coherent precession of the
spins of those muons for which a component of their spin polarization
lies perpendicular to this local field (expected to be 2/3 of the
total spin polarization for a powder sample). 
The frequency of the oscillations is given by
$\nu_{i} = \gamma_{\mu} |B_{i}|/2 \pi$, where $\gamma_{\mu}$ is the muon
gyromagnetic ratio ($=2 \pi \times 135.5$~MHz T$^{-1}$) and $B_{i}$
is the average magnitude of the local magnetic field at the $i$th muon
site. Any fluctuation in magnitude of these local fields will
result in a relaxation of the oscillating signal \cite{hayano}, described by
relaxation rates $\lambda_{i}$. 

Maximum entropy analysis (inset, Fig.~\ref{data}(c)) reveals two separate
frequencies in the spectra measured below $T_{\mathrm{C}}$,
corresponding to two magnetically inequivalent muon stopping sites in the
material. The precession frequencies, which are proportional to the internal
magnetic field experienced by the muon, 
 may be viewed as an effective order parameter
for these systems \cite{steve}. 
In order to extract the temperature dependence of the frequencies, 
the low temperature data were fitted to the function
\begin{eqnarray}
\label{fitlo}
A(t) &=&  \sum_{i=1}^{2} 
A_{i}\exp(-\lambda_{i} t)\cos(2 \pi \nu_{i} t)  \\ \nonumber
& & +A_{3}  \exp(-\lambda_{3} t) +A_{\mathrm{bg}},
\end{eqnarray}
where $A_{1}$ and $A_{2}$ are the amplitudes of the precession
signals and $A_{3}$ accounts for the contribution
from those muons with a spin component parallel to the local magnetic
field. The term $A_{\mathrm{bg}}$ reflects the
non-relaxing signal from those muons which stop in the sample holder
or cryostat tail.

The ratio of the two precession frequencies was found to be
$\nu_{2}/\nu_{1}=0.83$ across the temperature range $T <
T_{\mathrm{C}}$ and this ratio was fixed in the fitting procedure.
The amplitudes $A_{i}$ were found to be constant across the
temperature range and were fixed at 
values $A_{1}=1.66$\%, $A_{2}=3.74$\% 
and $A_{3}=5.54$\%. This shows that the probability of a muon
stopping in a site that gives rise to frequency $\nu_{1}$ is
approximately half that of a muon stopping in a site that corresponds
to $\nu_{2}$. We note also that $A_{3}$ is in excess of the expected
ratio of $A_{3}/(A_{1}+A_{2})=1/2$.
The unambiguous assignment of amplitudes is made
difficult by the resolution limitations that a pulsed muon source places on the
measurement. 
The initial muon pulse at ISIS has FWHM $\tau_{\mathrm{mp}} \sim 80$~ns,
 limiting the response for frequencies above
$\sim \tau_{\mathrm{mp}}^{-1}$ \cite{steve}. 
We should expect, therefore, slightly
reduced amplitudes or increased relaxation (see below) for the
oscillating components in our spectra for which $\nu_{1,2} \gtrsim 5$~MHz. 
The amplitudes of the oscillations are large enough, however, for us
to conclude that the
magnetic order in this material is an intrinsic property of the
bulk compound. 
Moreover, above $T_{\mathrm{C}}$ there is a complete recovery of the total
expected muon asymmetry. This observation, along with the constancy 
of $A_{1,2,3}$ below $T_{\mathrm{C}}$,
leads us to believe that Cs$_{2}$AgF$_{4}$ is completely
ordered throughout its bulk below $T_{\mathrm{C}}$.

\begin{figure}
\begin{center}
\epsfig{file=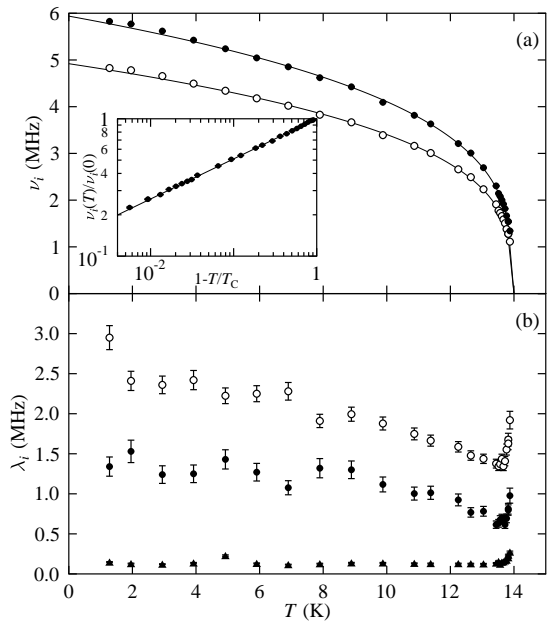,width=7.5cm}
\caption{Results of fitting data measured below $T_{\mathrm{C}}$
to Eq.~(\ref{fitlo}).
(a) Evolution of the muon-spin precession frequencies
  $\nu_{1}$ (closed circles) and $\nu_{2}$ (open circles) with
  temperature. 
Solid lines are fits to the
function $\nu_{i}(T) =\nu_{i}(0)(1-T/T_{\mathrm{C}})^\beta$ 
as described in
the text. {\it Inset:} Scaling plot of the precession frequencies 
with parameters 
$T_{\mathrm{C}}=13.95(3)$~K and $\beta=0.292$.
(b) Relaxation rates $\lambda_{1}$ (closed circles), $\lambda_{2}$ 
(open circles) and $\lambda_{3}$ (closed triangles),
as a function of temperature
showing a rapid increase as $T_{\mathrm{C}}$ is approached from below.
\label{fit}}
\end{center}
\end{figure}

Fig.~\ref{fit}(a) shows the evolution of the precession
frequencies $\nu_{i}$,  allowing us to investigate the
critical behavior associated with the phase transition. 
From fits of $\nu_{i}$ to the form 
$\nu_{i}(T) =\nu_{i}(0)(1-T/T_{\mathrm{C}})^{\beta}$ for
$T > 10$~K, 
we estimate $T_{\mathrm{C}}=13.95(3)$~K and $\beta =0.292(3)$. 
In fact, good fits to $\nu_{i}(T) =\nu_{i}(0)(1-T/13.95)^{0.292}$
are achieved over the entire measured temperature range (that is,
no spin wave related contribution was evident at low temperatures), yielding
$\nu_{1}(0)=6.0(1)$~MHz and $\nu_{2}(0)=4.9(2)$~MHz corresponding
to local magnetic fields at the two muon sites of 
$B_{1}=44(1)$~mT and $B_{2}=36(1)$~mT. 
A value of $\beta=0.292(3)$ is less than expected for three
dimensional  models
 ($\beta=0.367$ for 3D Heisenberg), but
larger than expected for 2D models ($\beta=0.23$ for 2D $XY$
or $\beta=0.125$ for
2D Ising) \cite{stevebook,bramwell}. This suggests that in
the critical regime
the behavior is intermediate in character between 2D and 3D; 
this contrasts with the magnetic properties
of  K$_{2}$CuF$_{4}$ where $\beta=0.33$, typical of
a 3D system, is observed 
in the reduced temperature region 
$t_{\mathrm{r}} \equiv (T_{\mathrm{C}}-T)/T_{\mathrm{C}} >7
\times 10^{-2}$, with a crossover to more 2D-like behavior
at $t_{\mathrm{r}} < 7\times 10^{-2}$, 
where $\beta=0.22$ \cite{hirakawa,hashimoto,suzuki}.
Our measurements probe the behavior of Cs$_{2}$AgF$_{4}$
for $t_{\mathrm{r}} \geq 5.5
\times 10^{-3}$, for which we do not observe any
crossover. 

A knowledge of $T_{\mathrm{C}}$ and the intraplane coupling $J$, allows us to
estimate the interplane coupling $J'$. Recent studies of
layered $S=1/2$ Heisenberg ferromagnets using the spin-rotation
invariant Green's function method \cite{schmalfuss}, show that
 the interlayer coupling may be described by an empirical formula
\begin{equation}
\left| \frac{J'}{J} \right| =  \exp \left( b - a \frac{|J|}{T_{\mathrm{C}}} \right)
\end{equation}
with $a=2.414$ and $b=2.506$. Substituting our value of 
$T_{\mathrm{C}}=13.95$~K and using $|J|/k_{\mathrm{B}}=44.0$~K \cite{mclain}, 
we obtain $|J'|/k_{\mathrm{B}}=0.266$~K and $|J'/J| =1.9 \times
10^{-2}$. The application of this procedure to K$_{2}$CuF$_{4}$
(for which $T_{\mathrm{C}}=6.25$~K and $|J|/k_{\mathrm{B}}=20.0$~K
\cite{yamada}) results in $|J'|/k_{\mathrm{B}}=0.078$~K and 
$|J'/J| = 3.9 \times
10^{-3}$. This suggests that, although 
highly anisotropic,
the interlayer coupling  is stronger in Cs$_{2}$AgF$_{4}$ than  
in K$_{2}$CuF$_{4}$. This may account for the lack of
dimensional crossover in Cs$_{2}$AgF$_{4}$ down to $t_{\mathrm{r}} = 5.5
\times 10^{-3}$.

Both transverse depolarization rates $\lambda_{1}$ and $\lambda_{2}$
are seen to decrease with increasing temperature (Fig.~\ref{fit}(b))
except close to $T_{\mathrm{C}}$ where they rapidly increase. 
The large values of $\lambda_{1,2}$ at low temperatures may reflect the reduced
frequency response of the signal due to the muon pulse width described
above. 
The large upturn in the depolarization rate close to $T_{\mathrm{C}}$,
which
is also seen in the longitudinal 
relaxation rate
$\lambda_{3}$ (which is small and nearly constant except on approach
to $T_{\mathrm{C}}$), may be attributed to the onset of critical fluctuations
close to $T_{\mathrm{C}}$. The component in the spectra with the
larger precession frequency $\nu_{1}$
has the smaller depolarization rate $\lambda_{1}$ at all
temperatures. These features provide further evidence for a magnetic phase
transition at $T_{\mathrm{C}}=13.95$~K.

Above $T_{\mathrm{C}}$ the character of the measured spectra changes
considerably (Fig.\ref{data}(a) and (b)) and we observe lower 
frequency oscillations
characteristic of the dipole-dipole interaction of the muon
and the $^{19}$F nucleus \cite{brewer}. 
 The
 Ag$^{2+}$ electronic moments, 
which dominate the spectra for $T<T_{\mathrm{C}}$,
 are no longer ordered in the paramagnetic regime,
and fluctuate very rapidly
on the muon time scale. They are therefore motionally narrowed
from the spectra, leaving the muon sensitive to the quasistatic nuclear 
magnetic moments. This interpretation is supported by $\mu^{+}$SR measurements
of K$_{2}$CuF$_{4}$ where similar behavior was observed \cite{mazzoli}.
In many materials containing
fluorine, the muon and two fluorine ions form a strong hydrogen
bond usually separated by approximately twice the F$^{-}$ ionic
radius. 
The linear F--$\mu^{+}$--F spin system consists of four distinct
energy levels with three allowed transitions between them 
(inset, Fig.~\ref{data}(b))
giving rise to the distinctive three-frequency oscillations observed. 
The signal is described by a
polarization function \cite{brewer} 
$
D(\omega_{\mathrm{d}}t) = \frac{1}{6} 
\left[ 3 + \sum_{j=1}^{3} u_{j} \cos (\omega_{j} t) \right],
$
where 
$u_{1}=1$, $u_{2}=(1+1/\sqrt{3})$ and $u_{3}=(1-1/\sqrt{3})$.
The transition frequencies (shown in Fig.~\ref{data}(b)) are given by 
$\omega_{j}=3 u_{j} \omega_{d}/2$
where $\omega_{d}=\mu_{0} \gamma_{\mu} \gamma_{\mathrm{F}}/4 \pi r^{3}$,
 $\gamma_{\mathrm{F}}$ is the $^{19}$F
nuclear gyromagnetic ratio and $r$ is the $\mu^{+}$--$^{19}$F separation.
This function accounts for the observed frequencies very well, leading us to
conclude that the F--$\mu^{+}$--F bonds are highly linear.

A successful fit of our data required the multiplication of 
$D(\omega_{\mathrm{d}} t)$ by
an exponential function with a small relaxation rate $\lambda_{4}$,
 crudely modelling fluctuations close to $T_{\mathrm{C}}$. The addition
of a further exponential component $A_{5}\exp(-\lambda_{5} t)$ was also
required in order to
account for those muon sites not strongly dipole coupled to fluorine nuclei. 
The data were fitted
with the resulting relaxation function
\begin{equation}
\label{eqhigh}
A(t) = A_{4} D(\omega_{d} t) \exp(-\lambda_{4} t) +A_{5}\exp(-\lambda_{5}
 t)  
+A_{\mathrm{bg}},
\end{equation}
The frequency $\omega_{\mathrm{d}}$ was found to be constant
at all measured temperatures, taking the value $\omega_{\mathrm{d}}=
2 \pi \times 0.211(1)$~MHz,
which corresponds to a constant F--$\mu^{+}$ separation of 1.19(1)~\AA, typical
of linear bonds \cite{brewer}. 
The relaxation rates only vary 
appreciably within 0.2~K of the magnetic transition, increasing as
$T_{\mathrm{C}}$ is approached from above, probably due to the
onset of critical fluctuations. This provides further evidence for our
assignment of $T_{\mathrm{C}}=13.95$~K.

Our determination of $\nu_{i}(0)$ and observation of the linear
F--$\mu^{+}$--F signal allow us to identify candidate
muon sites in Cs$_{2}$AgF$_{4}$. Although the magnetic structure of the
system is not known, magnetic measurements \cite{mclain} suggest
the existence of loosely coupled FM Ag$^{2+}$ layers
arranged antiferromagnetically along the $c$-direction.
Dipole fields were calculated for such a candidate magnetic structure
with Ag$^{2+}$ moments in the $ab$ planes oriented
parallel (antiparallel) to the $a$ direction for $z=0$ ($z=1/2$).
The calculation was limited to a sphere containing $\approx 10^5$ Ag ions with
localized moments of $0.8~\mu_{\mathrm{B}}$ \cite{mclain}.
The above considerations suggest that the muon sites will be situated
midway between two F$^{-}$ ions. 
Two sets of candidate muon sites may be identified in the planes
containing the fluorine ions. Magnetic fields corresponding to
$\nu_{2}(0)$ are found in the [CsF] planes (i.e.\
those with $z=0.145$ and $z=0.355$)
at the positions (1/4, 1/4, $z$), (1/4, 3/4, $z$), 
(3/4, 1/4, $z$) and (3/4, 3/4, $z$). 
Sites corresponding to the frequency $\nu_{1}(0)$
are more difficult to assign, but good candidates are found in the
[AgF$_{2}$] planes 
(at $z=0,1/2$) at positions (1/4, 1/2, $z$), 
(3/4, 1/2, $z$), (1/4, 0, $z$) and (3/4, 0, $z$). 
The candidate sites are shown in Fig.~\ref{structure}.
We note that there
are twice as many [CsF] planes in a unit
cell than there are [AgF$_{2}$] planes in agreement with our
observation that components with frequency $\nu_{2}$ occur with
twice the amplitude of those with $\nu_{1}$. 
Such an assignment then implies that the presence of the muon distorts 
the surrounding F$^{-}$ ions
such that their separation is $\sim 2.38$~\AA. This contrasts with
the in-plane F--F
separation in the unperturbed material of 4.55~\AA\ ([CsF]
planes)
and $\sim 3.2$~\AA\  ([AgF$_{2}$] planes) \cite{mclain}.
Thus the two adjacent F$^{-}$ ions in the magnetic [AgF$_{2}$]
planes each shift by $\sim 0.4$~\AA\ from their equilibrium
positions towards the $\mu^{+}$, demonstrating that the muon
introduces a non-negligible local distortion; however, the distortion
in the Ag$^{2+}$ ion positions is expected to be much less significant.

In conclusion, we have shown unambiguous evidence for magnetic order
in Cs$_{2}$AgF$_{4}$  with an exchange anisotropy of $|J'/J| \approx 10^{-2}$
and critical behavior intermediate in
character between 2D and 3D. The presence of coherent F--$\mu^{+}$--F
states allows a determination of candidate muon sites and an estimate
of the perturbation of the system caused by the muon probe. 
This study demonstrates that $\mu^{+}$SR is an effective and useful
probe of the Cs$_{2}$AgF$_{4}$ system. In order
to further explore this system as an analogue to the
high-$T_{\mathrm{C}}$ materials it is desirable to perform investigations
of doped materials based on the Cs$_{2}$AgF$_{4}$ parent compound.

Part of this work was carried out at the ISIS facility, Rutherford Appleton
Laboratory, UK. This work is supported by the EPSRC (UK).
T.L. acknowledges support from the Royal Commission for the Exhibition
of 1851. J.F.C.T and S.E.M acknowledge the U.S.\ National Science
Foundation under awards CAREER-CHE 039010 and OISE 0404938, respectively.

\end{document}